\documentclass[12pt,a4paper,final]{iopart}

\usepackage{iopams}  
\expandafter\let\csname equation*\endcsname\relax
\expandafter\let\csname endequation*\endcsname\relax
\usepackage{amsmath}
\usepackage{graphicx}

\usepackage[breaklinks=true,colorlinks=true,linkcolor=blue,urlcolor=blue,citecolor=blue]{hyperref}

\begin{document}

\title[]{Analytic solution to space-fractional Fokker-Planck equations for tempered-stable L\'{e}vy distributions with spatially linear, time-dependent drift }
\author{Mathew L. Zuparic}
\address{Defence Science and Technology Group, ACT 2600, Australia}
\ead{mathew.zuparic@dsto.defence.gov.au}
\author{Alexander C. Kalloniatis}
\address{Defence Science and Technology Group, ACT 2600, Australia}
\ead{alexander.kalloniatis@dsto.defence.gov.au}

\begin{abstract}
We derive analytic solutions for the full time dependence of space-fractional Fokker-Planck equations corresponding to stochastic Langevin equations with additive
tempered-stable L\'{e}vy noise terms. The drift terms are generalised to be spatially linear, but may contain arbitrary time dependence
such that no steady-state solution is available, even for the deterministic system.
\end{abstract}

\vspace{2pc}

\section{Introduction}
\label{intro}
The steady-state behaviour of many stochastic dynamical systems, while genuinely informative, is often the limit of what is studied by many 
researchers, particularly for non-Gaussian noise models. Analytic approaches are lacking when there is no steady-state behaviour, even deterministically. Here we solve the time-dependence of
Fokker-Planck (FP) probability densities in the presence of tempered stable noise. 

The classical method of solving both steady-state and time-dependent probability density functions for Fokker-Planck equations with additive Gaussian noise, the Ornstein-Uhlenbeck process, is covered in works such as Risken \cite{Risken89}. Time-dependence of some non-Gaussian FP equations, often based on multiplicative noise Langevin equations
(for example a noisy form of the logistic equation
\cite{Gora2005} drawing upon applications in financial mathematics by Linetsky \cite{Linetsky2004}), 
may be obtained by a variable transformation in a Gaussian system.
Here various orthogonal polynomials capture the space-dependence of the system; exponential decay weighted by
an eigenvalue of the Sturm-Liouville operator gives the time-dependence. This generalises the role of Hermite polynomials for the Ornstein-Uhlenbeck case  corresponding to additive Gaussian noise in the associated Langevin system.
Thus, various classical continuous orthogonal polynomials appear, leading to a mixed discrete-continuous spectrum for the time dependence \cite{Zuparic2015}. 
 
Moving beyond Gaussian based noise models is a growing area of interest. This is because many complex systems 
exhibit `fat-tail' phenomena \cite{BorovkovBorovkov08} which can lead to large spatial jumps, or large waiting times, such as in financial markets \cite{Kleinert2009,CarteaDelCast2007}, transport in plasmas \cite{delCastCarrLyn2005,delCastGonCheck2008} and
 brain activity \cite{RobBoonBreak2015}. 
The L{\'e}vy stable distribution
is a valuable model for the underlying stochastic system, where a power-law tail is modelled by an exponent $\alpha$. For $\alpha<2$ one obtains 
what is commonly known as fractional or anomalous diffusion \cite{MetzKlaft2000,MetzKlaft2004}. 
Here we find generalisations of the Ornstein-Uhlenbeck result where the Hermite polynomials in the spatial part of
the probability density function are replaced by Fox's H functions \cite{JMF1999}. Further generalisations of the external force, though limited to steady-state
solutions can be found in Chechkin \textit{et al.} \cite{CGKMT2002}. We focus in this paper on spatial jumps; on large waiting times 
see \cite{MetzKlaft2000b,MagWer2007,CGGKS2008,HenryLangStrak2010} or the bi-fractional case \cite{MetzNonn2002,Jumarie2004,KleinZat2013}.

A key property of stable L{\'e}vy noise is that because of the heavy tails there are no finite moments beyond the first for $1<\alpha<2$, and not
even the first for $0<\alpha<1$, a problem when fitting finite empirical datasets. 
Truncating with a hard cut-off \cite{Chambers76,MantStan1994} or, less
severely, tempering the noise \cite{Koponen1995,MeerSab2016} are two alternatives. The latter case is of interest to us here. 
A characteristic scale $\lambda$ moderates the extreme asymptotics of the tails rendering all moments finite. 
Thus far, only numerical \cite{BaeMeer2010,GajMag2010,KawMas2011} (the latter for time-fractional) or steady-state \cite{KullCast2012}
solutions of Fokker-Planck equations have been obtained for this case --- though the equations are solved up to a numerically computed inverse Fourier transform. We extend this approach to cover full time-dependence.
However, as a by-product we can generalise to the time-dependent solution the case of stable noise, where the Fourier-transform may be
explicitly computed. We focus on the spatially-tempered fractional Fokker-Planck equation but seek the solve as explicitly as possible for the entire
time-dependence, both for cases where steady-state behaviour is approached but also where the time-dependence may be
an ongoing regular dynamic, such as periodicity.

Our main motivation for obtaining analytic solutions to such generalised Ornstein-Uhlenbeck processes lies in their ability to provide insights and understanding into networked complex systems. Dynamical processes on networks remains an ongoing research area with far reaching physical, biological, social/organisational and chemical applications \cite{Dorfler2014}. The paradigmatic Kuramoto model of phase oscillators \cite{Kur84} offers an accessible mathematical formulation of networked dynamics which displays surprising emergent behaviour; for reviews refer to \cite{Acebron2005, Dorogovtsev2008,Arenas2008}. Our particular interest lies in understanding the model under stochastic influence. The deterministic system can be reduced to a system of decoupled first-order linear differential equations after linearisation close to synchronisation, where the system decays monotonically to pure frequency synchronisation \cite{Kallon2010}. With additive Gaussian noise the system becomes an Ornstein-Uhlenbeck system \cite{ZupKall2013}, and with additive tempered L\'{e}vy noise this is the tempered fractional Ornstein-Uhlenbeck system \cite{KallRob2017}; the drift in these cases is purely space-dependent.
Recently, the authors explored the cooperative and competitive behaviour displayed by the frustrated 2-network Kuramoto-Sakaguchi model \cite{Sakaguchi86}, both deterministically \cite{Kalloniatis2016} and under the influence of additive Gaussian noise \cite{Holder2017}. The relevant approximation for this system involves searching  states below the threshold for frequency synchronisation, where 2 or more internally synchronised populations may be in relative time-dependent oscillatory motion. In this case the drift terms of the corresponding Langevin equations (although spatially linear) are generally time-dependent. The latter work used a key result from Polyanin \cite{Polyanin02} which we show may be generalised to tempered stable noise, thus explaining at least one physics application of the results pursued in this work.

In the next section we set-up our formalism for the tempered-fractional Fokker-Planck equations, particularly using
the Fourier transform approach. In section \ref{SECAN} we then step through the solution, detailing both the stable case, and the more general tempered-stable case. In section \ref{SECNUM} we give some numerical examples which illustrate utility of this work by choosing drift terms which do not accommodate steady state density solutions. We also validate our analytic expression through numerical comparison with an alternate solution which is more computationally intensive (but more straightforwardly obtained). Finally in section \ref{SECCON} we offer conclusions and future applications of this work.

\section{Tempered-fractional-Fokker-Planck equations}
\subsection{Fractional Langevin equations}
We consider the fractional version of a Langevin equation
\begin{equation}
\dot{x} = q(x,t) + L^{\alpha,\theta,\lambda}(t),
\label{defining-Lang}
\end{equation}
where $L^{\alpha,\theta,\lambda}(t)$ is a tempered stable L{\'e}vy process in time described by parameters $\alpha,\theta,\lambda$. Here $\alpha \in (0,1) \cup (1,2]$ is the fractional power, governing the heavy-law tail of the distribution for $L$, $\theta \in [-1,1]$ is an asymmetry 
generating skew in the distribution, and $\lambda \in (0,\infty)$ is a tempering parameter that exponentially suppresses large jumps in the process $L$.
For $\alpha = 2$, the L\'{e}vy noise term in Eq. (\ref{defining-Lang}) becomes Gaussian. The case $\alpha=1$ is known as the \textit{Cauchy process}; we do not consider it in this work due to the `pathological' property of none of its moments formally existing \cite{Meerschaert01}. 

\subsection{Reimann-Liouville fractional derivatives}
The probability density $P(x,t)$ associated with the tempered-stable L\'{e}vy process given by Eq. (\ref{defining-Lang}) 
satisfies the Fokker-Planck equation \cite{Cartea07}
\begin{equation}
\frac{\partial}{\partial t} P (x,t) = \left( \Omega \partial^{\alpha,\theta,\lambda}_x - \frac{\partial}{\partial x} q(x,t)  \right)P(x,t), \;\; P(x,0) = \delta(x-y),
\label{defining-FP}
\end{equation}
where $\Omega \in (0,\infty)$ is the diffusion constant, which becomes the variance of the process in the Gaussian limit. Additionally, the operator $\partial^{\alpha,\theta,\lambda}_x$ is the tempered-fractional-diffusion operator, given explicitly as \cite{KullCast2012}
\begin{equation}
\partial^{\alpha,\theta,\lambda}_x = {\cal D}^{\alpha,\theta,\lambda}_x + v^{\alpha,\theta,\lambda} \frac{\partial}{\partial x} + \nu^{\alpha,\lambda},
\end{equation}
where $v^{\alpha,\theta,\lambda}$ and $\nu^{\alpha,\lambda}$ are additional drift and source/sink terms given by
\begin{equation}
v^{\alpha,\theta,\lambda}= \left\{  \begin{array}{cl}
0, & \alpha \in (0,1)\\
\frac{\alpha \theta \lambda^{\alpha-1}}{|\cos\frac{\pi \alpha}{2}|} & \alpha \in (1,2)
\end{array}
\right. , \;\; \nu^{\alpha,\lambda} =  \frac{\lambda^{\alpha}}{\cos\frac{\pi \alpha}{2}}.
\end{equation}
The operator ${\cal D}^{\alpha,\theta,\lambda}_x$ is called the $\lambda$-truncated fractional derivative of order $\alpha$, given by
\begin{equation}
{\cal D}^{\alpha,\theta,\lambda}_x = l(\theta) e^{-\lambda x} \,_{-\infty} D^{\alpha}_x e^{\lambda x} - r(\theta) e^{\lambda x} \,_x D^{\alpha}_{\infty} e^{-\lambda x},
\label{curlyD}
\end{equation}
where the operators $ \,_{-\infty} D^{\alpha}_x$ and $\,_x D^{\alpha}_{\infty}$ are the Riemann-Liouville derivatives defined as \cite{del-Castillo-Negrete12}
\begin{eqnarray}
\begin{split}
e^{-\lambda x} \,_{-\infty} D^{\alpha}_x e^{\lambda x} f(x) = \frac{e^{-\lambda x}}{\Gamma(m-\alpha)} \frac{\partial^m}{\partial x^m}\int^{x}_{-\infty}\frac{d\zeta e^{\lambda \zeta}}{(x-\zeta)^{\alpha+1-m}} f(\zeta) \\
e^{\lambda x}\,_x D^{\alpha}_{\infty}e^{-\lambda x}f(x) = \frac{(-1)^me^{\lambda x}}{\Gamma(m-\alpha)} \frac{\partial^m}{\partial x^m}\int^{\infty}_{x}\frac{d\zeta e^{-\lambda \zeta}}{(\zeta-x)^{\alpha+1-m}} f(\zeta),
\end{split}
\label{R-Lderiv}
\end{eqnarray}
for $\alpha-1< m < \alpha$. These can be equivalently defined in Fourier space \cite{Podlubny99,Samko93}
\begin{eqnarray}
\begin{split}
{\cal F} \left[e^{-\lambda x} \,_{-\infty} D^{\alpha}_x e^{\lambda x} f(x) \right] = (\lambda-i k)^{\alpha} \hat{f}(k)\\
{\cal F} \left[e^{\lambda x} \,_x D^{\alpha}_{\infty} e^{-\lambda x} f(x) \right] = (\lambda+i k)^{\alpha} \hat{f}(k),
\end{split}
\end{eqnarray}
where
\begin{eqnarray}
\begin{split}
{\cal F} \left[ f(x) \right] = \int^{\infty}_{-\infty}dx e^{i k x}f(x) =  \hat{f}(k)\\
{\cal F}^{-1} \left[ \hat{f}(k) \right] = \int^{\infty}_{-\infty} \frac{dk}{2\pi} e^{-i k x}\hat{f}(k) =  f(x).
\end{split}
\label{fourierdef}
\end{eqnarray}
The weighting factors
\begin{equation}
l(\theta) = \frac{\theta-1}{2 \cos\frac{\pi \alpha}{2}}, \;\; r(\theta) = \frac{\theta+1}{2 \cos\frac{\pi \alpha}{2}},
\end{equation}
determine the asymmetry imposed on each of the Riemann-Liouville derivatives. The introduction of the exponential decay terms to the Reimann-Liouville derivatives in Eqs. (\ref{curlyD}) and (\ref{R-Lderiv}) has the effect of \textit{tempering} the stable process. That is, for all $\alpha \in (0,1) \cup (1,2)$, \textit{all} moments of the corresponding tempered-stable process ($\lambda>0)$ are finite, marking a clear distinction from stable ($\lambda=0$) processes where only the first moment of the $\alpha \in (1,2)$ is finite.

\subsection{Spatially linear time-dependent drift terms}
We generalise the drift term $q(x,t)$ in Eqs. (\ref{defining-Lang}) and (\ref{defining-FP}) from the Ornstein-Uhlenbeck
form to include time-dependent coefficients
\begin{equation}
q(x,t) = \beta(t) - \gamma(t) x, \;\; \gamma(t) > 0,
\label{DRIFT}
\end{equation} 
where $\beta(t)$ and $\gamma(t)$ are general time-dependent functions. Hence, the explicit form of the tempered-fractional-Fokker-Planck equation (TFFP) becomes
\begin{equation}
\frac{\partial}{\partial t} P(x,t) = \left[ \Omega {\cal D}^{\alpha,\theta,\lambda}_x + \left( \gamma(t) x + \Omega  v^{\alpha,\theta,\lambda} - \beta(t)  \right) \frac{\partial}{\partial x}  + \left( \gamma(t) + \Omega \nu^{\alpha,\lambda}  \right) \right] P(x,t). 
\label{TFFP}
\end{equation}
For the Gaussian limit, i.e. $\partial^{\alpha,\theta,\lambda}_x = \frac{\partial^2}{\partial x^2}$, Polyanin \cite{Polyanin02} gives the necessary nonlinear transformations which result in the Gaussian equivalent of Eq. (\ref{TFFP}) becoming the standard heat equation. Indeed, this technique enabled the analytical investigation of clustering effects in the frustrated Kuramoto model under the influence of Gaussian white noise \cite{Holder2017}.  
In this work we follow a similar strategy and apply a variant of the nonlinear transformations given in \cite{Polyanin02} to Eq. (\ref{TFFP}) to obtain a tempered-fractional form of the heat equation with complications in the tempering. We then apply the Fourier transform to this `heat equation', solving it explicitly in Fourier space with the appropriate initial condition. The final solution is obtained by performing the inverse Fourier transform.

\section{Analytical solution to the tempered-fractional-Fokker-Planck equation}
\label{SECAN}
\subsection{Nonlinear transformations}
Generalising Polyanin \cite[section 1.8.3.6]{Polyanin02}  we consider the following set of transformations to ${ P}(x,t)$, $x$ and $t$:
\begin{eqnarray}
\begin{split}
{ P}(x,t) = { Q}(z,\tau) e^{\int^t_{0} d\xi \left(\gamma(\xi)+\Omega \nu^{\alpha,\lambda}  \right)},\;\; \tau =  \int^t_{t'} d\xi e^{\alpha \int^{\xi}_0 d\vartheta \gamma(\vartheta)}\\
z = x  e^{ \int^t_0 d\xi \gamma(\xi)} + \phi(t),\;\; \phi(t) = \int^t_0 d \xi e^{\int^{\xi}_0 d \vartheta\gamma (\vartheta)} \left( \Omega  v^{\alpha,\theta,\lambda} - \beta(\xi) \right),
\end{split}
\label{trans1}
\end{eqnarray}
for $t' \in \mathbb{R}$. Doing so results in the differential operator relations
\begin{eqnarray}
\begin{split}
\frac{\partial}{\partial t} =e^{\alpha \int^t_0 d\xi \gamma(\xi)} \frac{\partial}{\partial \tau} +e^{\int^t_0 d\xi \gamma(\xi)} \left(  \gamma(t) x + \Omega  v^{\alpha,\theta,\lambda} - \beta(t) \right) \frac{\partial}{\partial z}\\
 \frac{\partial}{\partial x} =e^{\int^t_0 d\xi \gamma(\xi)} \frac{\partial}{\partial z}.
\end{split}
\label{trans2}
\end{eqnarray}
Applying Eqs. (\ref{trans1}) and (\ref{trans2}) to Eq. (\ref{TFFP}) we obtain
\begin{equation}
e^{\alpha \int^t_0 d\xi \gamma(\xi)} \frac{\partial}{\partial \tau} { Q}(z,\tau) = \Omega {\cal D}^{\alpha,\theta,\lambda}_x   { Q}(z,\tau) ,
\label{heat-eq}
\end{equation}
with the initial condition $ { Q}(z,\tau(t= 0)) =\delta(z-y) $. We note that the $\lambda$-truncated fractional derivative of order $\alpha$ still contains the argument $x$, as opposed to $z$. This is removed by change of variables in the Riemann-Liouville derivatives of Eq. (\ref{R-Lderiv}):
\begin{eqnarray}
e^{-\lambda x} \,_{-\infty} D^{\alpha}_x e^{\lambda x} { Q}(z,\tau) = \frac{e^{-\lambda x}}{\Gamma(m-\alpha)} \frac{\partial^m}{\partial x^m}\int^{x}_{-\infty}\frac{d\zeta e^{\lambda \zeta} { Q}\left(\zeta e^{ \int^t_0 d\xi \gamma(\xi)} +\phi(t) ,\tau  \right)}{(x-\zeta)^{\alpha+1-m}} \nonumber\\
= e^{\alpha \int^t_0 d \xi \gamma(\xi) } \frac{e^{-\frac{\lambda}{e^{ \int^t_0 d\xi \gamma(\xi)}} z}}{\Gamma(m-\alpha)} \frac{\partial^m}{\partial z^m}\int^{z}_{-\infty}\frac{d\mu e^{\frac{\lambda}{e^{ \int^t_0 d\xi \gamma(\xi)}} \mu} { Q}\left(\mu ,\tau  \right)}{(z-\mu)^{\alpha+1-m}}\\
= e^{\alpha \int^t_0 d \xi \gamma(\xi) }  e^{-\frac{\lambda}{e^{ \int^t_0 d\xi \gamma(\xi)}} z} \,_{-\infty} D^{\alpha}_z e^{\frac{\lambda}{e^{ \int^t_0 d\xi \gamma(\xi)}} z} { Q}(z,\tau),\nonumber
\end{eqnarray}
where we have made the substitutions $x e^{ \int^t_0 d \xi \gamma(\xi) } = z - \phi(t)$ and $\zeta e^{ \int^t_0 d \xi \gamma(\xi) } =\mu-\phi(t)$ in the second line of the above expression. One can immediately show that an equivalent relation holds for the remaining Riemann-Liouville derivative, leading to
\begin{equation}
 {\cal D}^{\alpha,\theta,\lambda}_x   { Q}(z,\tau)  =  e^{\alpha \int^t_0 d \xi \gamma(\xi) }   {\cal D}^{\alpha,\theta,\frac{\lambda}{(d \tau/d t)^{1/\alpha}}}_z   { Q}(z,\tau) .
\end{equation} 
Hence, Eq. (\ref{heat-eq}) becomes
\begin{equation}
 \frac{\partial}{\partial \tau} { Q}(z,\tau) = \Omega {\cal D}^{\alpha,\theta,\frac{\lambda}{(d \tau/d t)^{1/\alpha}}}_z   { Q}(z,\tau)  .
\label{heat-eq2}
\end{equation}

Applying the Fourier transform Eq. (\ref{fourierdef}) to Eq. (\ref{heat-eq2}), one obtains an ordinary differential equation with respect to $\tau$ which has the solution
\begin{equation}
\hat{{ Q}}(k,\tau)=\kappa(k) \exp \left\{ \Omega  \left[  l(\theta)  \int d\tau \left(\frac{\lambda}{\left(\frac{d \tau}{d t}\right)^{1/\alpha}}-i k \right)^{\alpha} - r(\theta) \int d\tau  \left(\frac{\lambda}{\left(\frac{d \tau}{d t}\right)^{1/\alpha}}+i k \right)^{\alpha} \right] \right\},
\label{result1}
\end{equation}
where $ \kappa(k)$ is the constant of integration to be determined by the initial condition:
\begin{equation}
\hat{ Q}\left(k,\tau (t=0) \right) = e^{i k y}.
\end{equation}
We now solve Eq. (\ref{result1}) for the following two scenarios: 
\begin{itemize}
\item{$\lambda=0$ and general $\gamma(t)$, where we rely on the main results from \cite{Gorska11} to obtain an explicit expression for the inverse Fourier transform with time dependence;}
\item{$\lambda \ne 0$ and $\gamma(t) = \gamma \in \mathbb{R}$, where there is no known analytic expression for the inverse Fourier transform, 
so that we must rely on numerical integration in the final step.}
\end{itemize}

\subsection{Stable solution}
We consider first $\lambda=0$. Here we set $t' = 0$ in Eq. (\ref{trans1}) so that $\tau(t=0) = 0$. In this case, Eq. (\ref{result1}) becomes \cite{KallRob2017}
\begin{eqnarray}
\hat{{ Q}}(k,\tau)= e^{i k y}  \exp \left[ - \Omega \tau |k|^{\alpha} \left( 1 + i \theta \textrm{sgn}(k) \tan  \frac{\pi \alpha}{2} \right) \right].
\end{eqnarray}
Applying the inverse Fourier transform operation we obtain the L{\'e}vy-Khinchine formula,
\begin{eqnarray}
{ Q}(z,\tau) = \frac{1}{\pi \omega} \mathrm{Re} \int^{\infty}_0 dp  e^{-i p \frac{(z-y)}{\omega}} \exp \left( -p^{\alpha} e^{i \frac{\pi \chi}{2}} \right) ,
\label{gorska1}
\end{eqnarray}
where we have used the change of variables 
\begin{eqnarray}
\theta = \frac{\tan \frac{\pi \chi}{2} }{\tan \frac{\pi \alpha}{2}}, \;\; p = \omega k, \;\; \omega =  \left( \frac{\Omega \tau}{\cos \frac{\pi \alpha}{2}} \right)^{1/\alpha}.
\end{eqnarray}
If $\alpha$  and $\chi$ are rational numbers the inverse Fourier transform may be
explicitly evaluated. Specifically, for the integers $\{l,k,r\}$ we require,
\begin{equation}
\alpha = \frac{l}{k} , \;\; \chi = \frac{l - 2 r}{k} \textrm{ where } \left\{  \begin{array}{ccc} 0 <  \frac{l}{k} < 1, & 0 \le r \le l, & \textrm{ for $\alpha \in (0,1)$}\\
 1 <  \frac{l}{k} \le 2, & l-k \le r \le k,& \textrm{ for $\alpha \in (1,2]$}
\end{array} \right.,
\end{equation}
to obtain the following analytic expression for ${ Q}(z,\tau) $
\begin{equation}
{ Q}(z,\tau) = \frac{1}{\omega} \sum^{M-1}_{j=1}\frac{c_j(l,k,r)}{(\frac{z-y}{\omega})^{1\mp \frac{j l }{M}}}  \,_{m+1}F_M\left( \left. \begin{array}{c}
1, \; \Delta \left( m, \frac{1+jm}{M}\right)\\
\Delta\left( M,j+1 \right) \end{array}  \right| \frac{m^m (\frac{z-y}{\omega})^{\pm l}}{(-1)^{r-M}M^M} \right),
\label{resultlambda=0}
\end{equation}
where $M = \max(l,k)$, $m = \min(l,k)$, the upper and lower signs are used for the cases $\alpha \in (0,1)$ and $\alpha \in (1,2]$ respectively. Here, $\,_{m+1}F_M$ is the generalised hypergeometric function, $\Delta(i_1,i_2)$ is convenient notation for the parameter list,
\begin{eqnarray}
\Delta(i_1,i_2) = \left\{ \frac{i_2}{i_1}, \frac{i_2 +1}{i_1}, \dots, \frac{i_1+i_2-1}{i_1} \right\},
\end{eqnarray}
and the coefficients $c_j(l,k,r)$ are given by,
\begin{equation}
c_j(l,k,r) = \frac{M^{\frac{1}{2} -j}m^{\frac{1}{2} + \frac{j m}{M}}}{2^{-r}(2\pi)^{\frac{l+k}{2}}}\frac{\left[ \prod^{j}_{i=1} \Gamma \left( \frac{i-j-1}{M} \right) \right] \left[ \prod^{M}_{i=j+2} \Gamma \left( \frac{i-j-1}{M} \right) \right]}{\left[ \prod^{m-1}_{i=0} \Gamma \left( \frac{j}{M}+ \frac{i+1}{m} \right) \right]^{-1}\left[ \prod^{r-1}_{i=0} \sin \left( \pi \left\{ \frac{i}{r}+ \frac{j}{M}\right\} \right) \right]^{-1}}.
\end{equation}
Eq. (\ref{resultlambda=0}) is given as the main result of Gorska and Penson \cite{Gorska11} which involves the application of the Mellin transform and subsequent Meijer-G function identities to Eq. (\ref{gorska1}).

Thus, with Eqs. (\ref{trans1}) and (\ref{resultlambda=0}) with rational fractions for $\alpha$ and $\chi$, the density ${ P}(x,t)$ for the $\lambda=0$ case is
\begin{eqnarray}
{ P}(x,t)= { Q}\left( x  e^{ \int^t_0 d\xi \gamma(\xi)} + \phi(t),\int^t_{0} d\xi e^{\alpha \int^{\xi}_0 d\vartheta \gamma(\vartheta)} \right) e^{\int^t_{0} d\xi \gamma(\xi)}.
\end{eqnarray}
For irrational  $\alpha$ and $\chi$ in the stable $\lambda=0$ case, numerical integration is still required for the inverse Fourier transform.

\subsection{Tempered-stable solution}
Turning to $\lambda\neq 0$, we set $t' = -\infty$ in Eq. (\ref{trans1}) so that $\tau(t=0) = 1/(\alpha \gamma)$. In order to proceed we set $\gamma(t) = \gamma \in \mathbb{R}_+$, which leads to
\begin{eqnarray}
\left(\frac{\lambda}{\left(\frac{d \tau}{d t}\right)^{1/\alpha}}\mp i k \right)^{\alpha} =  \left(\frac{\varphi}{\tau^{1/\alpha}} \mp i k \right)^{\alpha} \;\; \textrm{where} \;\; \varphi = \frac{\lambda}{(\alpha \gamma)^{1/\alpha}}.
\end{eqnarray}
The integrals can then be evaluated using the hypergeometric expressions 
\begin{eqnarray}
\begin{split}
\int d \tau \left(\frac{\varphi}{\tau^{1/\alpha}} \mp i k \right)^{\alpha} = (\mp i k )^{\alpha} \tau  \,_2F_1\left( \left. \begin{array}{cc}
-\alpha,& - \alpha\\
& 1-\alpha \end{array}  \right| \mp \frac{i \varphi}{\tau^{1/\alpha} k} \right)\\
=  \left(  \varphi \mp i \tau^{1/\alpha} k \right)^{\alpha} \,_2F_1\left( \left. \begin{array}{cc}
-\alpha,& 1\\
& 1-\alpha \end{array}  \right| \frac{\varphi}{\varphi \mp i \tau^{1/\alpha} k }\right)\\
= { Y}_{\mp}(k,\tau),
\end{split}
\label{2F1tech}
\end{eqnarray}
where we have used Abramowitz and Stegun \cite[Eq. (15.3.4)]{Abramowitz72} to generate the second line of Eq. (\ref{2F1tech}). Hence the Fourier transform of the TFFP density is
\begin{eqnarray}
\hat{{ Q}}(k,\tau) = \kappa(k) \exp \left\{ \Omega \left[  l(\theta) { Y}_{-}(k,\tau) - r(\theta)  { Y}_{+}(k,\tau) \right] \right\},
\end{eqnarray}
where the integration constant $\kappa(k)$ is given by
\begin{eqnarray}
\kappa(k) = e^{iky} \exp \left\{ -\Omega \left[  l(\theta) { Y}_{-}\left(k,\frac{1}{\alpha \gamma} \right) - r(\theta) { Y}_{+}\left(k,\frac{1}{\alpha \gamma} \right) \right] \right\}.
\end{eqnarray}
Additionally, applying the convenient notation
\begin{eqnarray}
{ Z}(k,\tau) = { Y}_{-}\left(k,\tau \right)-{ Y}_{-}\left(k,\frac{1}{\alpha \gamma} \right),
\end{eqnarray}
we obtain
\begin{equation}
\hat{{ Q}}(k,\tau) = e^{iky} \exp \left\{ -\frac{\Omega}{ \cos\frac{\pi \alpha}{2}} \left[ \mathrm{Re}( { Z}(k,\tau) ) - i \theta  \mathrm{Im} ({ Z}(k,\tau) )  \right]\right\}.
\label{finalFourier}
\end{equation}
Hence  with Eqs. (\ref{trans1}) and (\ref{finalFourier}) the solution for the density ${\cal P}(x,t)$ becomes
\begin{eqnarray}
{ P}(x,t) = e^{\left(\gamma+\Omega \nu^{\alpha,\lambda}  \right)t} \mathrm{Re} \int^{\infty}_{0}\frac{d k}{\pi} e^{- i k( x e^{\gamma t} + \phi(t))} \hat { Q}\left(k,\frac{1}{\alpha \gamma} e^{\alpha \gamma t} \right),
\label{finalP}
\end{eqnarray}
where we have reduced the problem of computing the probability density to a one-dimensional integral.

\section{Numerical examples and validation}
\label{SECNUM}
\subsection{Damped oscillatory drift}
\begin{figure}
\begin{center}
  \includegraphics[width=0.9\linewidth]{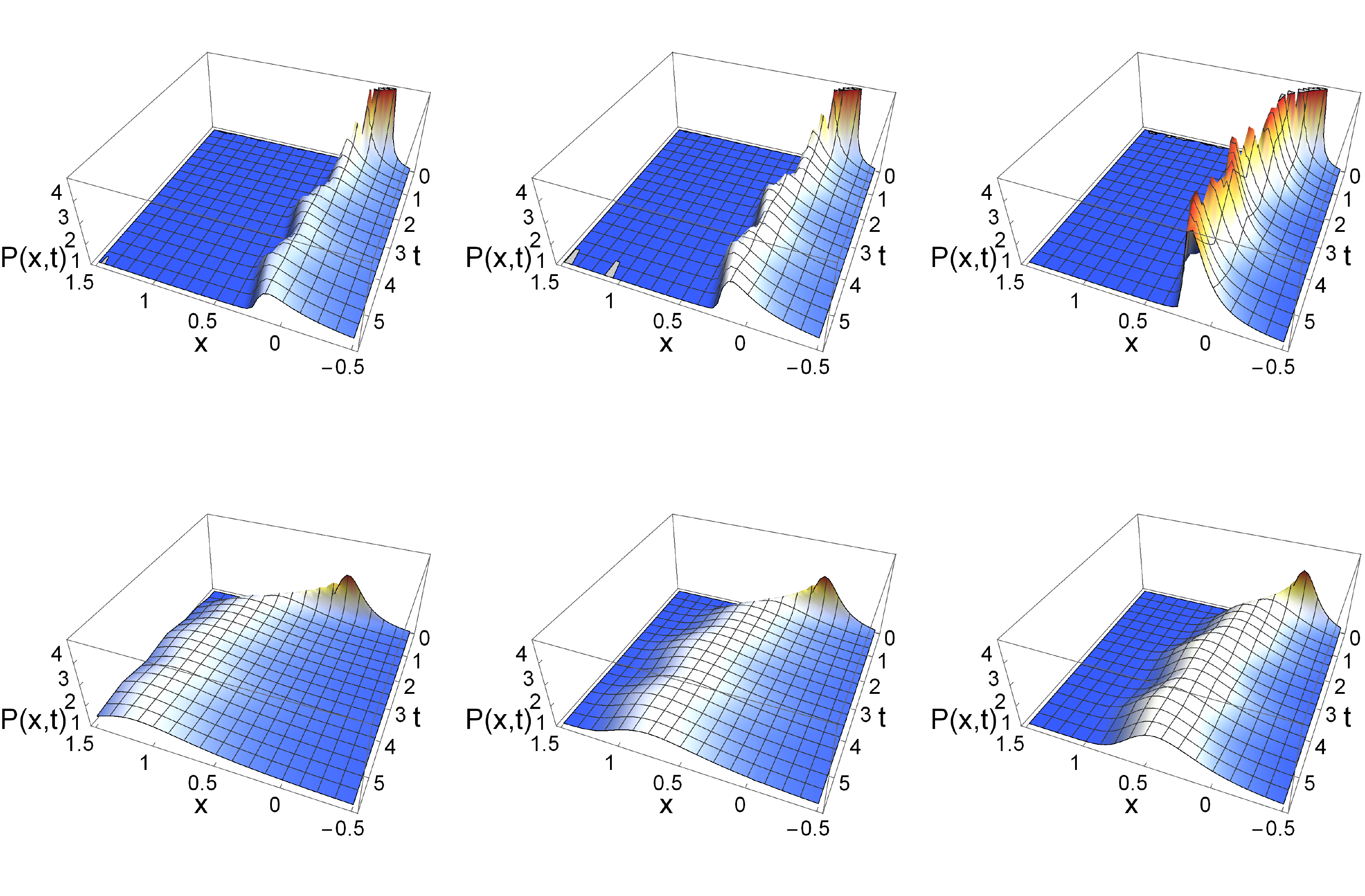}
  \caption{Plots of the time-dependent probability density function ${P}(x,t)$ using Eq. (\ref{finalP}) with
    $\beta(t)=1-e^{-t}+\frac{1}{4}\sin \pi t$, $\theta=0.9$, $y=-0.3$, $\Omega=1$, $\gamma=3$. Top row: $\alpha=0.55$, left $\lambda=0.001$, middle $\lambda=0.05$, right $\lambda = 0.9$. Bottom row: $\alpha=1.15$, left $\lambda=0.001$, middle $\lambda = 0.05$, right $\lambda = 0.9$. \label{fig:densityprof}}
\end{center}
\end{figure}

To illustrate the time-dependence of the solution we choose the drift to have the form
\begin{eqnarray}
\beta(t)=1-e^{-t}+\frac{1}{4}\sin \pi t .
\end{eqnarray}
This incorporates elements of exponential decay, but not to a constant. Contrastingly we choose a constant for the coefficient of $x$ in $q(x,t)$,
specifically $\gamma=3$. We fix the noise strength $\Omega=1$ and asymmetry $\theta=0.9$. We plot ${ P}(x,t)$ for a range of $\alpha$ above and below $\alpha=1$, and increasing $\lambda$, through applying the NIntegrate function in Mathematica\textsuperscript{\textregistered} 10.4, using the AdaptiveQuasiMonteCarlo method. In figure \ref{fig:densityprof} we show plots for $\alpha=0.55$ (top row) and $1.15$ (bottom row) with $\lambda=0.001, 0.05$ and $0.9$
(left to right). Additionally we remark that each of the six panels' first time step is $t=0.1$.

We observe for increasing $\lambda$ the peaking of the density and, concurrently, exposing the time-dependence. Specifically, we see the decay from
the initial condition down to the oscillatory behaviour. For both values of $\alpha$ we observe the asymmetry and the heavy-tail in the negative
direction. The key difference between $\alpha<1$ and $\alpha>1$ is also evident: the plots in the top row show the cusp like behaviour characteristic of
the former case whereas the lower row is smoother, more closely approaching Gaussian shapes. The most significant difference between the two
cases is the induced drift in the $\alpha<1$ case. We note that at large $t$, $\beta(t)$ oscillates about unity. In the
probability density for $\alpha<1$ this oscillation is shifted in the negative direction. For $\alpha>1$, the density oscillates
closer to what may be expected deterministically.

\subsection{Numerical validation}

In order to validate Eq. (\ref{finalP}), and the plots provided in figure \ref{fig:densityprof}, we compare outputs from Eq. (\ref{finalP}) to a more direct and well known method of solution to Eq. (\ref{defining-FP}): the method of characteristics \cite{Evans10}. Through comparing outputs to a more elementary method of solution (though with the downside of being more computationally intensive) we are able to ensure that Eq. (\ref{finalP}) is indeed correct. 

To begin we perform a Fourier transform on Eq. (\ref{defining-FP}) to obatin the first order partial differential equation
\begin{eqnarray}
\begin{split}
\left(\frac{\partial}{\partial t} + \gamma k \frac{\partial}{\partial k}\right) \hat{{P}}(k,t) = \left( \Omega \Lambda(k) + i k \beta (t) \right)\hat{{P}}(k,t), \;\; \hat{{P}}(k,0) = e^{iky},\\
\textrm{where  }\;\; \Lambda(k) = l(\theta) (\lambda - ik)^{\alpha} - r(\theta) (\lambda + ik)^{\alpha} - i k v^{\alpha,\theta,\lambda} + \nu^{\alpha,\lambda}.
\end{split}
\label{defining-FP2}
\end{eqnarray}
Following Evans  \cite[Chapter 3]{Evans10}, Eq. (\ref{defining-FP2}) can be recast as following characteristic ordinary differential Initial Value Problems (IVPs)
\begin{eqnarray}
\begin{split}
\frac{d}{d\xi} t(\mu,\xi) = 1, &\;\; t (\mu, 0) = 0,\\
\frac{d}{d\xi} k(\mu,\xi) = \gamma k, &\;\; k(\mu,0) = \mu, \\
\frac{d}{d\xi} \hat{{P}}(\mu,\xi) = \left(\Omega \Lambda(k) + i k \beta (t)\right)\hat{{P}} , &\;\; \hat{{P}}(\mu,0) = e^{i \mu y} ,
\end{split}
\label{defining-FP3}
\end{eqnarray}
for parametric variables $\mu$ and $\xi$. 

\begin{figure}
\begin{center}
  \includegraphics[width=0.9\linewidth]{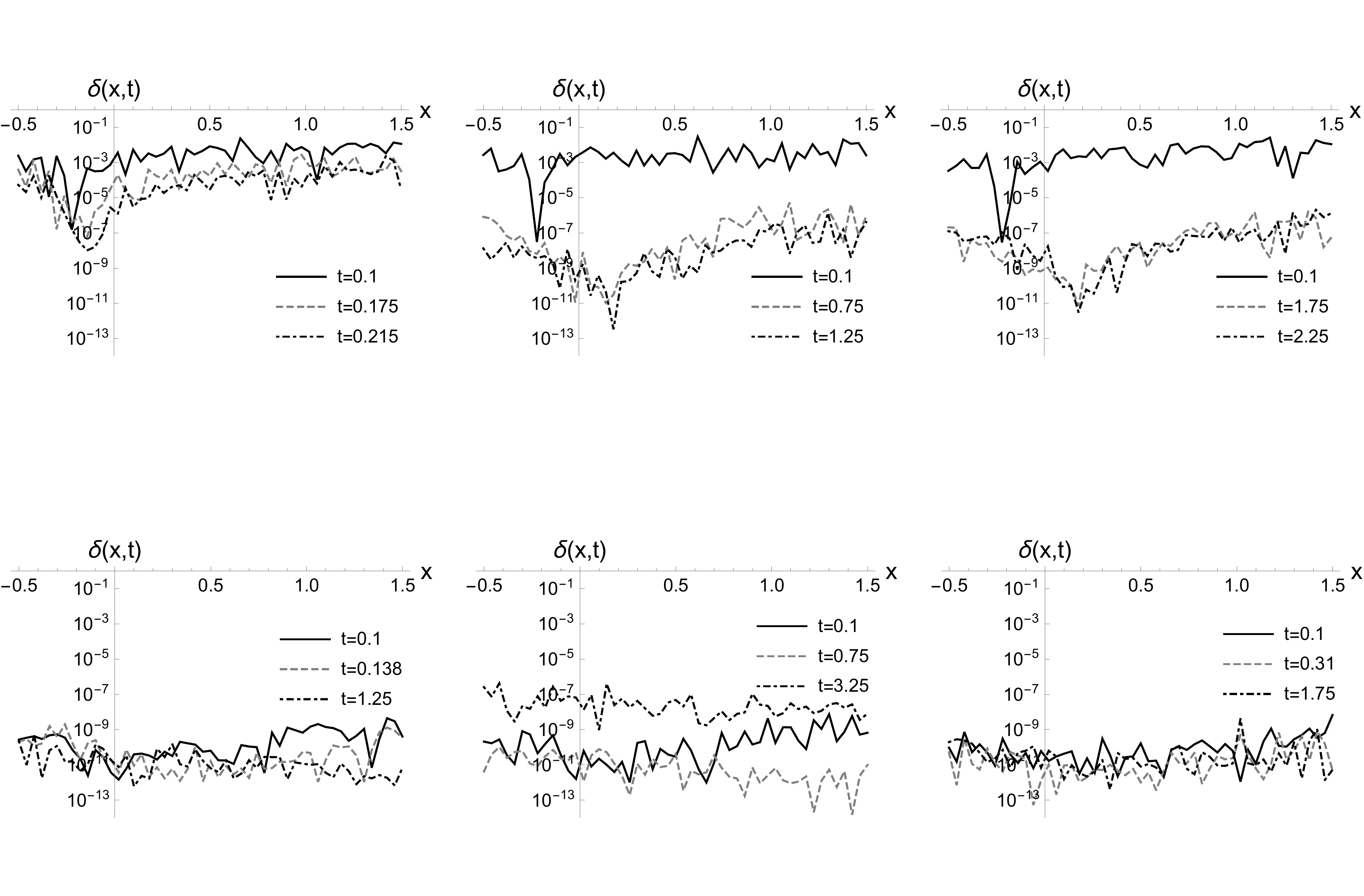}
  \caption{Plots for $\delta(x,t)$: logarithmic plots of the modulus of difference between the probability density function calculated through Eq. (\ref{finalval}), and Eq. (\ref{finalP}). Parameter choices for each plot are equivalent to figure \ref{fig:densityprof}: $\beta(t)=1-e^{-t}+\frac{1}{4}\sin \pi t$, $\theta=0.9$, $y=-0.3$, $\Omega=1$, $\gamma=3$ common for all. Top row: $\alpha=0.55$, left $\lambda=0.001$, middle $\lambda=0.05$, right $\lambda = 0.9$. Bottom row: $\alpha=1.15$, left $\lambda=0.001$, middle $\lambda = 0.05$, right $\lambda = 0.9$. \label{fig:delta}}
\end{center}
\end{figure}

Solving the first two IVPs in Eq. (\ref{defining-FP3}) reveals
\begin{equation}
\begin{array}{l}
 t(\mu,\xi) = \xi\\
  k(\mu,\xi) = \mu e^{\gamma \xi} 
\end{array}  \Rightarrow 
\begin{array}{ll}
\xi (k, t)= t\\
\mu (k, t)= k  e^{-\gamma t}
\end{array}.
\end{equation}
Hence, the solution to the third IVP in Eq. (\ref{defining-FP3}) is
\begin{eqnarray}
\begin{split}
& \hat{{P}}(\mu,\xi) = e^{i \mu y} \exp \left[ \int^{\xi}_{0}dq \left( \Omega \Lambda \left(\mu e^{\gamma q} \right) + i \mu e^{\gamma q} \beta (q) \right) \right]\\
 \Rightarrow \;\;&  \hat{{P}}(k,t) = e^{i  k  e^{-\gamma t} y} \exp \left[ \int^{t}_{0}dq \left( \Omega \Lambda \left(k e^{\gamma (q-t)} \right) + i k e^{\gamma (q-t)} \beta (q) \right) \right].
 \end{split}
 \label{defining-FP4}
\end{eqnarray}
To the best of our knowledge there is no analytic form to the integral in Eq. (\ref{defining-FP4}), equivalent to the result of Eq. (\ref{finalFourier}) as a solution to Eq. (\ref{result1}). Thus, in order to calculate the probability density, we are required to perform 2 integrals in the form,
\begin{equation}
{P}(x,t) = \frac{1}{2 \pi} \int^{\infty}_{-\infty}dk e^{-i k x} \hat{{P}}(k,t) ,
\label{finalval}
\end{equation}
for $ \hat{{P}}(k,t) $ given by Eq. (\ref{defining-FP4}).

\begin{figure}
\begin{center}
  \includegraphics[width=0.4\linewidth]{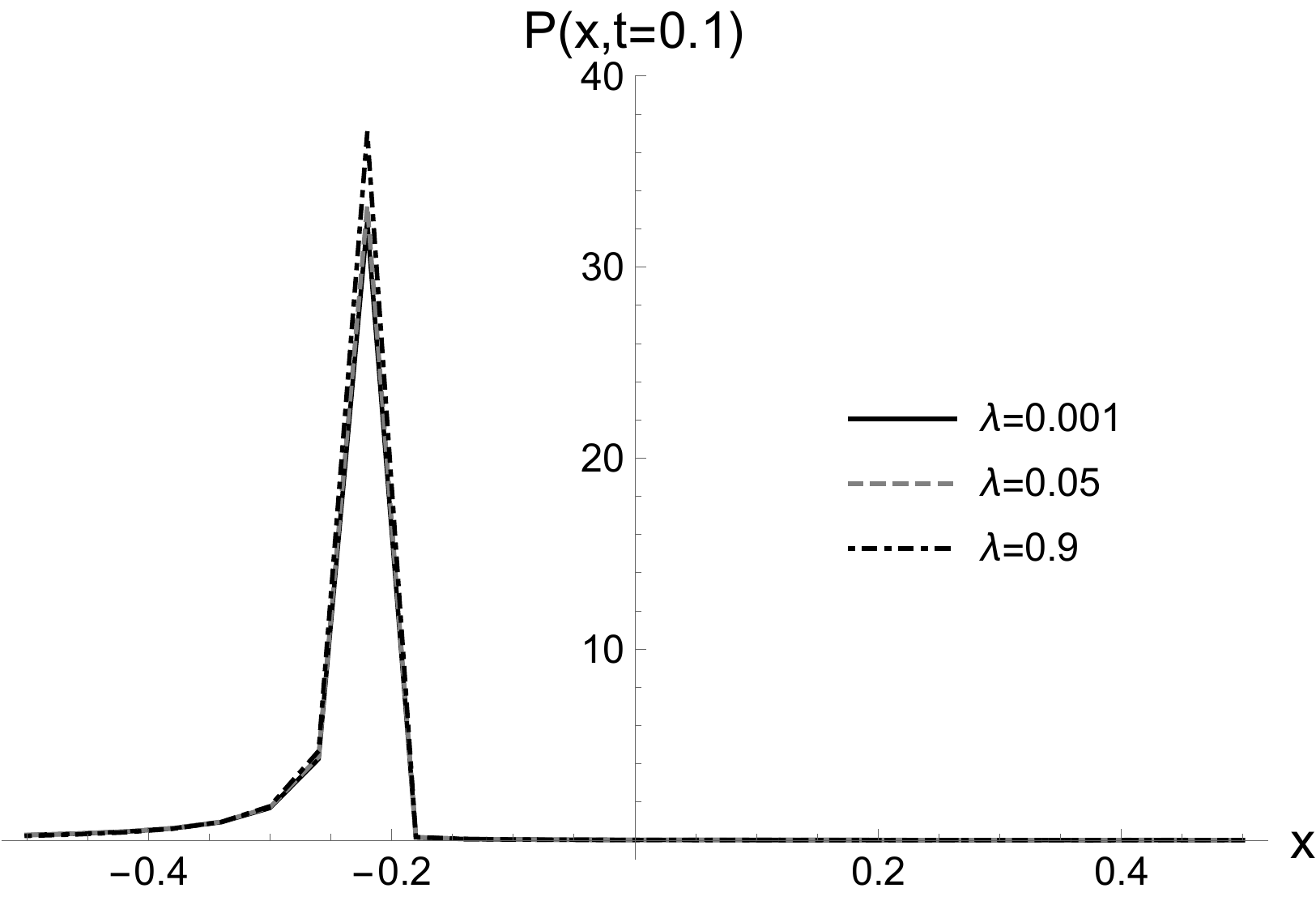}
  \caption{Plots for $P(x,0.1)$ generated from Eq. (\ref{finalP}). Parameter choices for each plot are $\beta(t)=1-e^{-t}+\frac{1}{4}\sin \pi t$, $\theta=0.9$, $y=-0.3$, $\Omega=1$, $\gamma=3$ and $\alpha=0.55$. \label{fig:t01}}
\end{center}
\end{figure}

In figure \ref{fig:delta} (generated using the same software and method as figure \ref{fig:densityprof}) we give the logarithmic plot of the modulus of the difference between the numerical calculations of Eq. (\ref{finalval}), and Eq. (\ref{finalP}) (labeled as $\delta(x,t)$) for various time instances. Each of the six panels corresponds to the same parameter choices given in figure \ref{fig:densityprof}, given explicitly in the caption. Visual inspection of figure \ref{fig:delta} shows that the difference between calculating $P(x,t)$ numerically through either Eq. (\ref{finalP}) or Eq. (\ref{finalval}) is quite small, rarely exceeding $10^{-2}$. Macroscopically, we notice that the values for $\delta(x,t)$ on the bottom panels are generally much lower than those for the top panels, especially for small $t$. A likely explanation for this observation is the more `delta-function-like' behaviour of the densities for $\alpha = 0.55$ at $t=0.1$. Indeed, we notice that the remaining curves on the top left panel, for $t=\{0.175,0.215\}$, display greater values of $\delta(x,t)$ than the curves for $t=\{0.75,1.25\}$ and $t=\{1.75,2.25\}$ for the top-middle and top-right panels respectively. Referring back to the top panels in figure \ref{fig:densityprof}, we see that the densities are quite highly peaked for small times, only losing this behaviour at approximately $t > 0.7$, thus explaining why the corresponding $\delta(x,t)$ plots in the top-middle and top-right panels are much lower. Almost counter-intuitively however, the sharp `dips' in the top panels for the $t=0.1$ plots correspond to the peaks of the densities for this particular time, all of which rise higher than a value of $30$. The corresponding plots of $P(x,0.1)$ for $\alpha=0.55$ and various choices of $\lambda$ are given in figure \ref{fig:t01}.

The plots given in the bottom panels of figure \ref{fig:delta} contain no surprises. The corresponding density plots in figure \ref{fig:densityprof} for $\alpha = 1.15$ clearly show much more diffuse behaviour for $t \ge 0.1$ than those for $\alpha = 0.55$. Correspondingly, the values of $\delta(x,t)$ are much smaller for these parameter choices, rarely rising above $10^{-7}$, and achieving values as low as $10^{-13}$. Thus, through the comparison of Eq. (\ref{finalP}) with Eq. (\ref{finalval}) over a range of parameter values, we conclude that the analytical result offered by Eq. (\ref{finalP}) is indeed valid.

\section{Conclusions and future work}
\label{SECCON}
We have solved the tempered fractional Fokker-Planck equation with spatially linear drift and time-dependent coefficients,
giving the full time-dependence of the probability density function up to a single integral that must be numerically integrated. 
For the stable noise case but with rational fractional parameters even the final inverse Fourier integral may be analytically computed. Our approach exploits a rarely used
nonlinear transformation to absorb the space-dependence revealing a tempered-fractional heat equation easily exploited by the Fourier transform.
The solution invokes a range of hypergeometric functions associated with the time integral. We have validated our expression through comparing outputs with another solution technique
involving a relatively straightforward application of the method of characteristics and then numerical integration. 

As previously mentioned, our intention is to apply the results contained in this work to stochastic versions of the Kuramoto model \cite{Kur84}
of synchronising oscillators on networks. The model subject to L{\'e}vy noise has been addressed in
\cite{KallRob2017}, where approximations in the vicinity of complete phase synchronisation are applied.
Going beyond this involves approximating in the vicinity of partial synchronisation, where oscillators may coalesce into
two clusters that may or may not be locked with respect to each other, or where two populations of oscillators are interacting. 
This problem maps precisely to the form of drift considered in this paper
and has been solved for Gaussian noise in \cite{Holder2017}.
We are thus positioned to solve the tempered fractional stochastic generalisation of this as well as a broader set of multi-population
models.

\section*{Acknowledgements}
We are grateful for discussions with Dale Roberts. ACK was supported by a Chief Defence Scientist Fellowship.

\end{document}